\documentclass[journal,onecolumn]{IEEEtran}

\usepackage{ifthen}
\usepackage{theorem}
\usepackage{graphicx}
\usepackage{psfrag}
\usepackage{amsfonts}
\usepackage[cmex10]{amsmath}
\usepackage{cite}
\usepackage{url}

\def\tcbr{\ifCLASSOPTIONtwocolumn\\\fi}
\def\ochf{\ifCLASSOPTIONtwocolumn\else\hfill\fi}

\newtheorem{Thm}{Theorem}

\def\half{\frac{1}{2}}

\newcommand{\ud}{\,{\mathrm d}}

\newcommand{\vect}[2]{#1_1,#1_2,\ldots,#1_#2}
\newcommand{\vectsup}[2]{#1^1,#1^2,\ldots,#1^#2}

\newcommand{\bv}[1]{\mathbf{#1}}

\def\eps{\epsilon}

\def\tran{\mathsf{T}}

\def\cqsc{C_{\text{$q$-SC}}}
\def\cqscs{C_{\text{$q$-SC*}}}
\def\lchqsc{L_{\text{ch}}}
\def\qsc{$q$-SC}
\def\qscs{$q$-SC*}

\def\bvynj{\mathbf{Y}^{[j]}=\mathbf{y}^{[j]}}
\def\pji{p^j_i}
\def\pjk{p^j_k}

\def\epsbsc{\epsilon_{\text{BSC}}}
\def\cbsc{C_{\text{BSC}}}

\newcommand{\msg}[3]{\mu_{#1 \to #2}(#3)}
\newcommand{\marg}[1]{\sum_{\sim\{#1\}}}

\def\xj{x^j}
\def\xji{x^j_i}
\def\yj{y^j}
\def\yji{y^j_i}
\def\ej{e^j}
\def\eji{e^j_i}

\def\Xj{X^j}
\def\Xji{X^j_i}
\def\Yj{Y^j}
\def\Yji{Y^j_i}

\def\Eji{E^j_i}

\def\ferr{f_E}
\def\fch{f_\text{ch}}
\def\fchj{f_\text{ch}^j}

\allowdisplaybreaks[1]

\begin{document}

\title{A Fresh Look at Coding for\\ $q$-ary Symmetric Channels}

\author{Claudio Weidmann and Gottfried Lechner%
  \thanks{The material in this paper was presented in part at the IEEE
    International Symposium on Information Theory, Toronto, Canada, Jul.\
    2008, and at the 5th International Symposium on Turbo Codes \& Related
    Topics, Lausanne, Switzerland, Sep.\ 2008. Parts of this work were
    conducted while the authors were with ftw.\ Telecommunications Research
    Center Vienna, Austria.}%
  \thanks{Claudio Weidmann is with ETIS -- CNRS UMR 8051 / ENSEA / Univ
    Cergy-Pontoise, France.}%
  \thanks{Gottfried Lechner is with the Institute for Telecommunications
    Research, University of South Australia, Mawson Lakes Boulevard, Mawson
    Lakes 5095, South Australia, Australia. This work has been supported by
    the Australian Research Council under ARC grant DP0881160.}%
  \thanks{E-mail: claudio.weidmann@ieee.org, gottfried.lechner@unisa.edu.au}%
}

\maketitle 

\begin{abstract}
  This paper studies coding schemes for the $q$-ary symmetric channel based on
  binary low-density parity-check (LDPC) codes that work for any alphabet size
  $q=2^m$, $m\in\mathbb{N}$, thus complementing some recently proposed
  packet-based schemes requiring large $q$.  First, theoretical optimality of
  a simple layered scheme is shown, then a practical coding scheme based on a
  simple modification of standard binary LDPC decoding is proposed. The
  decoder is derived from first principles and using a factor-graph
  representation of a front-end that maps $q$-ary symbols to groups of $m$
  bits connected to a binary code. The front-end can be processed with a
  complexity that is linear in $m=\log_2 q$. An extrinsic information transfer
  chart analysis is carried out and used for code optimization. Finally, it is
  shown how the same decoder structure can also be applied to a larger class
  of $q$-ary channels.
\end{abstract}


\section{Introduction}
\label{sec:intro}

\IEEEPARstart{T}{he} $q$-ary symmetric channel ($q$-SC) with error probability
$\eps$ takes a $q$-ary symbol at its input and outputs either the unchanged
input symbol, with probability $1-\eps$, or one of the other $q-1$ symbols,
with probability $\eps/(q-1)$. It has attracted some attention recently as a
more general channel model for packet-based error correction.  For very large
$q$, its appropriately normalized capacity approaches that of an erasure
(packet loss) channel.  In the following, we will only consider channel
alphabets of size $q=2^m$ with $m \in \mathbb{N}$.

The capacity of the $q$-SC with error probability $\eps$ is 
\begin{equation*}
  \cqsc = m - h(\eps) - \eps\log_2(2^m-1) 
\end{equation*}
bits per channel use, where $h(x)=-x\log_2 x - (1\!-\!x)\log_2(1\!-\!x)$ is
the binary entropy function.  Asymptotically in $m$, the normalized capacity
$\cqsc/m$ thus approaches $1\!-\!\eps$, which is the capacity of the binary
erasure channel (BEC) with erasure probability $\eps$.

\begin{figure}[b]
\centering
\includegraphics[width=3.2in]{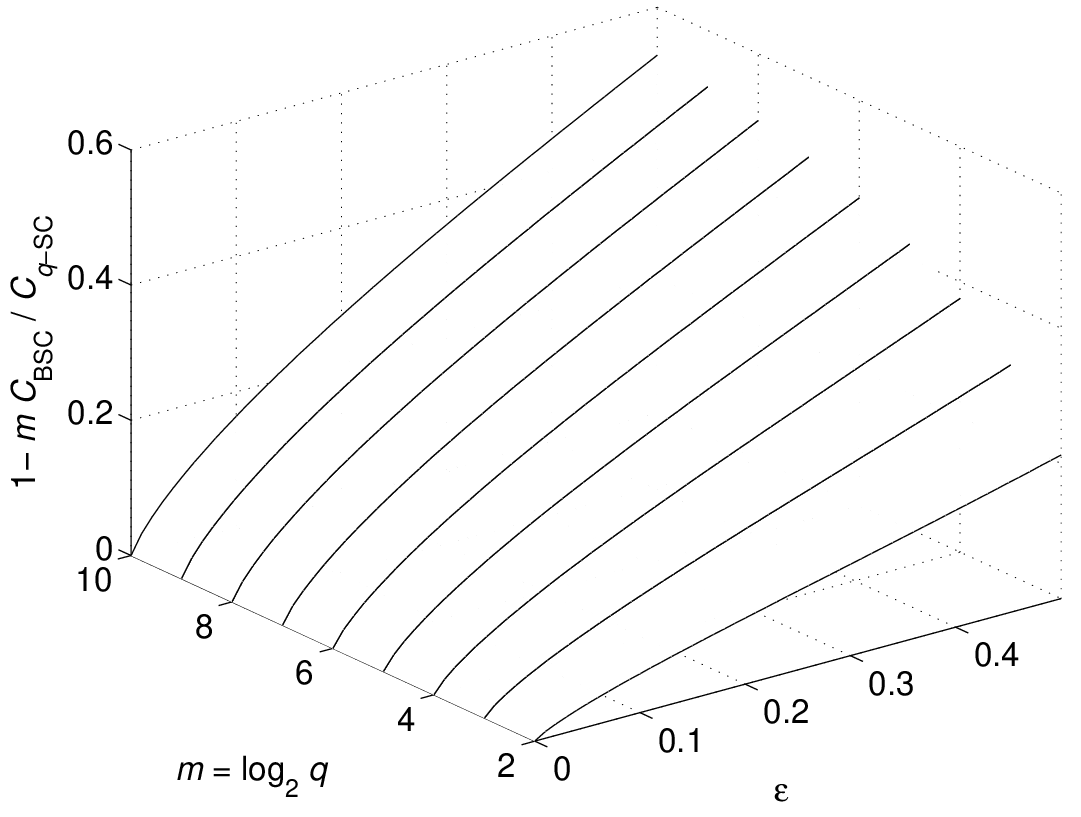}
\caption{Relative capacity loss $1- \frac{m\cbsc}{\cqsc}$ of marginal BSC vs.\
  $q$-SC.}
\label{fig:qsc_vs_bsc_3d}
\end{figure}

\begin{figure}[t]
\centering
\includegraphics[width=3.2in]{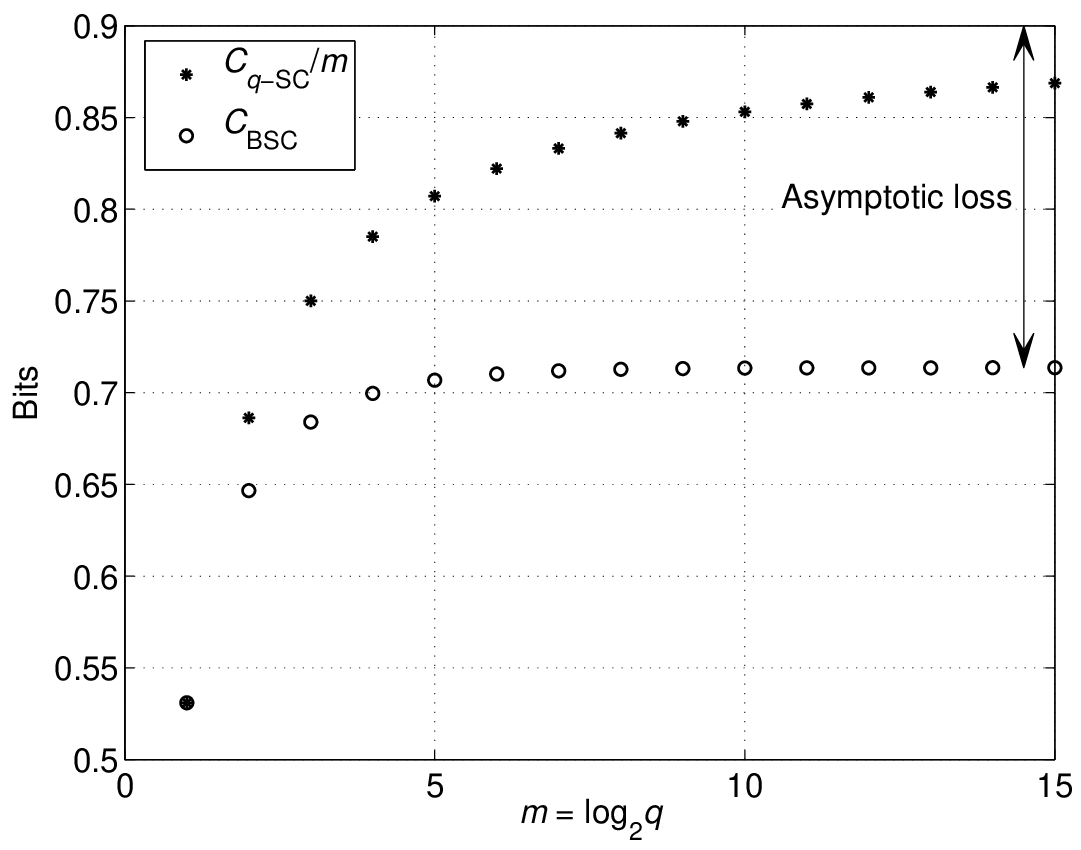}
\caption{Normalized capacity of $q$-SC vs.\ marginal BSC (error probability
  $\eps=0.1$).}
\label{fig:qsc_vs_bsc}
\end{figure}

Recent work
\cite{Luby2005,Brown2004,Shokrollahi2004a,Shokrollahi2004b,Zhang2011} has
shown that it is possible to approach $\cqsc$ for large alphabet sizes
$q=2^m$, with symbols of hundreds to thousands of bits, and complexity $O(\log
q)$ per code symbol. The focus of the present work is on smaller $q$, with
symbols of tens of bits at most, although the presented coding techniques will
work for any $q=2^m$.

The $q$-ary channel input and output symbols will be represented by binary
vectors of length $m$.
Hence a simplistic coding approach consists in decomposing the $q$-SC into $m$
binary symmetric channels (BSCs) with crossover probability 
\begin{equation}
  \label{eq:eps_bsc}
  \epsbsc = \frac{q \eps}{2(q-1)} =\frac{\eps}{2(1-2^{-m})},
\end{equation}
which have capacity $\cbsc=1-h(\epsbsc)$ each. 

We briefly study the normalized capacity loss, $\Delta = \cqsc/m-\cbsc$, which
results from (wrongly) assuming that the $q$-SC is composed of independent
BSCs. For fixed $m$, we have $\lim_{\eps\to 0} \Delta =0$; so using binary
codes with independent decoders on the $m$-fold BSC decomposition might be
good enough for small $\eps$ (e.g.\ $\eps<10^{-3}$). However,
Figure~\ref{fig:qsc_vs_bsc_3d} shows that the relative capacity loss
${m\Delta}/{\cqsc}=1-{m\cbsc}/{\cqsc}$ increases close to linearly in $\eps$.
For fixed $\eps$, we have $\lim_{m\to \infty} \Delta = h(\eps/2)-\eps$, which
can be a substantial fraction of the normalized $q$-SC capacity (e.g., for
$\eps=0.1$, $h(\eps/2)-\eps=0.19$).  Figure~\ref{fig:qsc_vs_bsc} shows that
already for small $m$, the $q$-SC capacity is substantially larger than what
can be achieved with the BSC decomposition.  Clearly, there is a need for
coding schemes targeted at ``large'' $\eps$ (say $\eps>10^{-1}$), but moderate
$m$ (say $2 \leq m < 20$), which are not that well handled by methods for
large $q$. For example, for using verification-based decoding the symbols
should have between $m=32$ bits for short codes (block length $10^2\ldots
10^3$) and $m=64$ bits for longer codes (block length $10^4\ldots 10^5$)
\cite{Luby2005}.

One example application, which also motivated this work, is Slepian-Wolf
coding of $q$-ary sources with a discrete impulse-noise correlation model,
using $q$-SC channel code syndromes as fixed-rate (block) source code.  This
can then be used as a building block for a Wyner-Ziv coding scheme, where $q$
is the number of scalar quantizer levels, or for fixed-rate quantization of
sparse signals with a Bernoulli-$\eps$ prior on being nonzero
\cite{Weidmann2008b}. Clearly, $q$ will be only moderately large in such a
scenario.

The capacity loss of the binary decomposition is due to the fact that the
correlation between errors on the binary sub-channels is not taken into
account, i.e., since an error on one sub-channel of the \qsc\ implies a symbol
error, the error probabilities on the other sub-channels will change
conditioned on this event.  A better approach would be the use of non-binary
low-density parity-check (LDPC) codes over GF($q$) \cite{Davey1998}. While
that would take into account the dependency between bit errors within a
symbol, the decoding complexity of the associated non-binary LDPC decoder is
$O(q\cdot\log q)$ or at least $O(q)$ when using sub-optimal algorithms
\cite{Declercq2007}.

Instead, this work focuses on a modified binary LDPC decoder of complexity
$O(\log q)$. Section~\ref{sec:layer_scheme} studies an ideal scheme using
layers of different-rate binary codes, providing the key intuition that once a
bit error is detected, the remaining bits of the symbol may be treated as
erasures without loss in rate.  Section~\ref{sec:bit-APP} then proposes a
scheme using a single binary code and develops the new variable node decoding
rules from first principles, by factorizing the posterior
probabilities. Section~\ref{sec:front-end} shows that the new decoding rule is
equivalent to a front-end (or \emph{first receiver block}) that maps $q$-ary
symbols to groups of $m$ bits and studies its factor graph representation.
Section~\ref{sec:exit} analyzes the extrinsic information transfer chart
characterization of this \qsc\ front-end, which is shown to allow
capacity-achieving iterative decoders. Section~\ref{sec:design} briefly
outlines how to design optimized LDPC codes for this problem and presents
simulation results. Finally, Section~\ref{sec:gen_scheme} extends these
decoding methods to a larger class of $q$-ary channels.

\section{Layered Coding Scheme}
\label{sec:layer_scheme}

We study the following layered coding scheme based on binary codes. Blocks of
$k$ symbols $[\vectsup{u}{k}]$ are split into $m$ bit layers
$[u_i^1,u_i^2,\ldots,u_i^k]$, $i=1,\ldots,m$, and each layer is independently
encoded with a code for a binary symmetric erasure channel (BSEC) with erasure
probability $\delta_i$ and crossover probability $\epsilon_i$, to be specified
below.  The channel input symbol will be denoted $\xj=[\vect{\xj}{m}]^\tran$,
where $\xji$ is the $i$-th bit of the $j$-th symbol, while the corresponding
channel output is $\yj=[\vect{\yj}{m}]^\tran$, respectively $\yji$.
 
The key idea is to decode the layers in a fixed order and to declare \emph{bit
  erasures} at those symbol positions in which a bit error occurred in a
previously decoded layer. This saves on the code redundancy needed in the
later layers, since erasures can be corrected with less redundancy than bit
errors. 

It will be shown that this layered scheme is optimal if the constituent BSEC
codes attain capacity. The intuition behind the optimality of this seemingly
suboptimal scheme is that once a bit error (and thus a symbol error) has been
detected, all the following layers have bit error probability 1/2 in that
position, since the $q$-SC assigns uniform probabilities over the possible
symbol error values. Now the BSC(1/2) has zero capacity and so the concerned
bits can be treated as erasures with no loss.

The decoder performs successive decoding of the $m$ layers, starting from
layer 1. All errors corrected at layer $i$ and below are forwarded to layer
$i+1$ as erasures, that is all bit error positions found in layers 1 up to $i$
will be marked as erased in layer $i+1$, even though the channel provides a
(possibly correct) binary output for those positions.  Let $\epsilon_i$ be the
probability that the channel output $y$ is equal to the input $x$ in bit
positions 1 to $i-1$ and differs in position $i$ (i.e.\
$[\vect{y}{{i-1}}]^\tran = [\vect{x}{{i-1}}]^\tran$ and $y_i \neq x_i$). A
simple counting argument shows that there are $2^{m-i}$ such binary vectors
$y \neq x$, out of a total $2^m-1$. The $i$-th binary sub-channel is thus
characterized by
\begin{align}
\label{eq:eps}
  \epsilon_i &= \frac{2^{m-i}}{2^m-1}\,\eps,\\
\label{eq:delta}
  \delta_i &= \sum_{j=1}^{i-1} \epsilon_j = \frac{2^m-2^{m-i+1}}{2^m-1}\,\eps. 
\end{align}

\begin{Thm}
  The layered scheme achieves $q$-SC capacity if the constituent BSEC codes
  achieve capacity.
\end{Thm}

\begin{IEEEproof}
  We may assume an ideal scheme, in which all layers operate at their
  respective BSEC capacities and correct all errors and erasures. The
  BSEC($\delta_i,\epsilon_i$) capacity is
  \begin{equation}
    \label{eq:bsec}
    C_{\mathrm{BSEC}} =
    (1-\delta_i)\left(1-h\left(\frac{\epsilon_i}{1-\delta_i}\right)\right). 
  \end{equation}
  Hence the sum of the layer rates becomes
  \begin{align}
\label{eq:pf11}
    \sum_{i=1}^m R_i &= \sum_{i=1}^m 
    (1-\delta_i)\left(1-h\left(\frac{\epsilon_i}{1-\delta_i}\right)\right)\\
\label{eq:pf12}
&= m  +  \sum_{i=1}^m \bigg\{ -\delta_i
-(1-\delta_i)\log_2(1-\delta_i)
\ifthenelse{\boolean{CLASSOPTIONtwocolumn}}{\nonumber\\ &  \qquad}{}
+ \epsilon_i\log_2\epsilon_i + (1-\delta_{i+1})\log_2(1-\delta_{i+1})
\bigg\}\\
\label{eq:pf13}
&= m + \sum_{i=1}^m \big\{ -(m-i)\epsilon_i +\epsilon_i\log_2\epsilon_i \big\}
\ifthenelse{\boolean{CLASSOPTIONtwocolumn}}{\nonumber\\ &  \qquad}{}
+(1-p)\log_2(1-p)\\
\label{eq:pf14}
&= m + \sum_{i=1}^m \big\{ -(m-i)\epsilon_i + (m-i)\epsilon_i \big\}
+p\log_2 p
\ifthenelse{\boolean{CLASSOPTIONtwocolumn}}{\nonumber\\ &  \qquad}{}
-p\log_2(2^m-1) + (1-p)\log_2(1-p)\\
&= m - h(p) - p\log_2(2^m-1) = \cqsc,\nonumber
  \end{align}
where \eqref{eq:pf11} follows from \eqref{eq:bsec} and the definition of the
layered scheme, \eqref{eq:pf12} follows from $ \delta_i + \epsilon_i =
\delta_{i+1}$ (which holds up to $i=m$, when $\delta_{m+1}=\eps$),
\eqref{eq:pf13} follows from the evaluation of the telescoping sum and
$\sum_{i=1}^m\delta_i = \sum_{i=1}^m (m-i)\epsilon_i$, and \eqref{eq:pf14}
follows from substituting \eqref{eq:eps} for $\epsilon_i$.
\end{IEEEproof}

As can be seen from inserting \eqref{eq:eps} and \eqref{eq:delta} into
\eqref{eq:bsec}, the layer rate quickly approaches $1-\eps$ for increasing
$i$, that is, for large $m$ the last layers all operate at rates close to
$1-\eps$. Thus an interesting variant of the layered scheme is to use $\mu<m$
BSEC layers as above, followed by a single ``thicker'' layer, which sends the
remaining $m-\mu$ bits (per symbol) over the same
BSEC($\delta_{\mu+1},\epsilon_{\mu+1}$). In particular, this could even be
beneficial in practical implementations, since combining layers leads to
longer codewords and thus better codes, which might outweigh the theoretical
rate loss.  The next section will show that it is actually possible to reap
the benefits of a single large binary code, without any layering, by using a
decoder that exploits the dependencies among the bits in a symbol.

The layered scheme has several disadvantages compared to the symmetric scheme
presented below. The main disadvantage is that in practical implementations,
each layer will need a different BSEC code that is tuned to the effective
erasure and error probabilities resulting from the layers preceding
it. Besides causing problems with error propagation, this is also impractical
for hardware implementation, since the required silicon area would necessarily
grow with the number of layers. Another disadvantage is that for a fixed
symbol block length $n$, the layered scheme uses binary codes of length $n$,
while the symmetric scheme uses block length $mn$. The latter will provide a
performance advantage, especially for short block lengths.

\section{Bit-Symmetric Coding Scheme}
\label{sec:sym_scheme}

Ideally, a coding scheme for the $q$-SC should be symmetric in the bits
composing a symbol, that is, no artificial hierarchy among bit layers should
be introduced (notice that the order of the bit layers may be chosen
arbitrarily). We propose to encode all bits composing the symbols with one
``big'' binary code, which needs to satisfy just slightly stricter constraints
than an ordinary code, while the decoder alone will exploit the knowledge
about the underlying $q$-SC. The key concept that should carry over from the
layered scheme is that the decoder is able to declare erasures at certain
symbol positions and thus needs less error correction capability (for part of
the bits in erased symbols).

We present two approaches to describe the decoder for the proposed scheme: a
bit-level APP approach, which clearly displays the probabilities being
estimated (and is more intuitive to generalize, see
Sec.~\ref{sec:gen_scheme}), and a decoder front-end approach, which
facilitates the use of EXIT chart tools.

\subsection{Bit-Level APP Approach}
\label{sec:bit-APP}

Our proposal for a practical symmetric $q$-SC coding scheme relies on an LDPC
code with information block length $K\!=\!m k$ bits and channel block length
$N\!=\!m n$ bits.  We assume that the variable nodes (VNs) in the decoder
receive independent extrinsic soft estimates of $\Xji$ (that is, bit $i$ of
code symbol/vector $\Xj$, for $i=1,\ldots,m$, $j=1,\ldots, n$) from the check
nodes (CNs). These amount to estimates of $P(\Xji|\bv{Y}^{[j]}=\bv{y}^{[j]})$
or the corresponding $\log$-likelihood ratio (LLR), $L(X) =
\log(\Pr(X\!=\!0)/\Pr(X\!=\!1))$.  (As usual, the notation $\bv{y}^{[j]}$
denotes the block consisting of all symbols/vectors except the $j$-th.)  In
particular, the extrinsic estimate of $\Xji$ is assumed to be independent of
the other bits $X^j_{[i]}$ of the same symbol, so that we may write
$$
P(\Xji|X^j_{1},\ldots,X^j_{i-1},
\bv{Y}^{[j]}\!=\!\bv{y}^{[j]})=P(\Xji|\bv{Y}^{[j]}\!=\!\bv{y}^{[j]}).
$$ 

In the standard case, these independence assumptions are justified by the fact
that asymptotically in the block length, the neighborhood of a VN in the LDPC
decoder computation graph becomes a tree
\cite[Chap.~3]{MCT2007}. Unfortunately, the $q$-SC VN message computation rule
has to depend on the other bits in the same symbol, in order to account for
the bit error correlation, and thus will introduce cycles. However, this
problem can be alleviated by imposing an additional constraint on the code,
namely that the parity checks containing $\Xji$ do not involve any of the bits
$X^j_{[i]}$. This is a necessary condition for the above intra-symbol
independence assumption and may be achieved by using an appropriate edge
interleaver in the LDPC construction. Then the cycles introduced by the VN
message rule will grow asymptotically and are thus not expected to lead to
problems in practice.

As suggested above, the properties of the $q$-SC can be taken into account via
a simple modification of the VN computation in the message-passing decoding
algorithm for binary LDPC codes.  We factor the \emph{a posteriori}
probability (APP) of symbol $\Xj$ as follows:
\begin{align}
  P(\Xj|\bv{Y}=\bv{y}) &\doteq 
P(\Yj=\yj|\Xj) P(\Xj|\bvynj)\nonumber\\
\label{eq:sap2}
&= P(\Yj=\yj|\Xj) \prod_{i=1}^m
P(\Xji|\bvynj), 
\end{align}
where the factorization in \eqref{eq:sap2} is made possible by the above
independence assumption (the symbol $\doteq$ denotes equality up to a positive
normalization constant). Using the definition of the $q$-SC, this becomes
\begin{multline}
  \label{eq:sap3}
\ochf  P(\Xj=\xj|\bv{Y}=\bv{y}) \tcbr \doteq
  \begin{cases}
    (1-\eps)\prod_{i=1}^m
P(\Xji=\xji|\bvynj), & \xj = \yj,\\
    \frac{\eps}{q-1}\prod_{i=1}^m
P(\Xji=\xji|\bvynj), & \xj \neq \yj.
  \end{cases}
\ochf
\end{multline}
We define the extrinsic probability that $\Xj = \yj$ as
\begin{equation*}
  \beta^j = \prod_{i=1}^m P(\Xji=\yji|\bvynj)
= \prod_{i=1}^m \pji,
\end{equation*}
where we introduced
\begin{equation*}
  \pji=P(\Xji\!=\!\yji|\bvynj)
\end{equation*}
for notational convenience. The normalization
constant in \eqref{eq:sap3} thus becomes
\begin{equation*}
  \gamma^j = (1-\eps)\beta^j + \frac{\eps}{q-1} (1-\beta^j).
\end{equation*}
Then the bit APP may be obtained by the marginalization
\begin{equation}
  \label{eq:bap1}
  P(\Xji=\xji|\bv{Y}=\bv{y}) 
= \sum_{x^j_{[i]} \in \{0,1\}^{m-1}} P(\Xj=\xj|\bv{Y}=\bv{y}),
\end{equation}
which may be written as
\begin{multline}
  \label{eq:bap2}
\ochf P(\Xji=\xji|\bv{Y}=\bv{y}) \tcbr 
=  \begin{cases}
  \left[(1-\eps)\beta^j_{[i]}+\frac{\eps}{q-1}(1-\beta^j_{[i]})
      \right] \cdot \frac{\pji}{\gamma^j}, 
      & \xji = \yji,\\
      \frac{\eps}{q-1}
      \cdot \frac{1-\pji}{\gamma^j},
      & \xji \neq \yji,
  \end{cases} 
\ochf
\end{multline}
where $\beta^j_{[i]} \!=\! \beta^j / \pji$
is the intra-symbol
extrinsic probability that $\Xj = \yj$, using no information on bit
$\Xji$. Finally, we may express the 
\emph{a posteriori} bit-level LLR as
\begin{align}
\label{eq:appllr}
\ifthenelse{\boolean{CLASSOPTIONtwocolumn}}{&}{}%
L_{\mathrm{app}}(\Xji\!=\!\yji) 
\ifthenelse{\boolean{CLASSOPTIONtwocolumn}}{\nonumber\\ }{}%
&= \log\left( 
    \frac{(q-1)\beta^j-q \eps\beta^j+\eps \pji}{\eps(1-\pji)} 
  \right) \nonumber\\
 &= \log\left( 
    \frac{(q-1)\beta^j_{[i]}-q \eps\beta^j_{[i]}+\eps }{\eps}
    \cdot\frac{\pji}{1-\pji} 
  \right) \nonumber\\
  &= \underbrace{\log\left( 1+ \frac{q-q \eps -1}{\eps}\cdot\beta^j_{[i]}
\right)}_{\text{\normalsize $\lchqsc(\Xji\!=\!\yji)$}}
+ L_{\mathrm{extr}}(\Xji\!=\!\yji).
\end{align}
The usual $L(X)$ is obtained from $L(X\!=\!y) =
\log(\Pr(X\!=\!y)/\Pr(X\!=\!1-y))$ via a sign flip, $L(X)=(1-2y)L(X\!=\!y)$.

The second term in \eqref{eq:appllr} corresponds to the extrinsic information
from the CNs that is processed at the VNs in order to compute the bit APP in
standard binary LDPC decoding. The difference lies in the first term in
\eqref{eq:appllr}, which corresponds to the channel LLR (which would be
$L_{\mathrm{ch}}(X)=\log(P(y|X\!=\!0)/P(y|X\!=\!1))$ in the binary case). When
the extrinsic information on the bits $X^j_{[i]}$ favors the hypothesis
$\Xj \neq \yj$, the product $\beta^j_{[i]}$ will be small and therefore
$\lchqsc$ in \eqref{eq:appllr} will be close to zero, which is equivalent to
declaring a bit erasure. This shows that the symmetric LDPC scheme relies on
``distributed'' \emph{soft} bit erasure estimates, while in the layered scheme
the erasures are declared in a \emph{hard} ``top-down'' fashion.

Equation \eqref{eq:appllr} describes the modification of the VN computation
that turns a message-passing binary LDPC decoder into one for the $q$-ary
symmetric channel. The outgoing VN messages are computed as usual by
subtracting the incoming edge message from $L_{\mathrm{app}}(\Xji)$; also the
CN messages are the same as in the binary case.  For practical implementation
purposes, \eqref{eq:appllr} should probably be modified (approximated) in
order to avoid switching back and forth between probabilities and LLRs when
computing $\beta^j$. A final detail is the specification of the initial
channel LLR $\lchqsc^{(0)}$ in \eqref{eq:appllr}, which is needed to start the
decoder iterations. By inserting the memoryless worst-case estimate
$\beta^j_{[i]}=2^{-m+1}$ into \eqref{eq:appllr}, we obtain
\begin{equation}
  \label{eq:startLch}
   \lchqsc^{(0)} =  \log\left(\frac{2(1-2^{-m})-\eps}{\eps} \right),
\end{equation}
which is exactly the channel LLR for the marginal BSC with crossover
probability $\epsbsc$ given in \eqref{eq:eps_bsc}.

Notice that the decoder iterations are exclusively between VN
\eqref{eq:appllr} and CN computations, like in the binary case. However,
computing \eqref{eq:appllr} at the VNs requires the extrinsic information (the
CN messages) for all bits within a symbol; this might be considered an
additional level of message exchanges (specifically, plain copying of
messages), but it does not involve iterations of any kind. The complexity
increase compared to binary LDPC decoding is on the order of at most $m$
operations per variable node, depending on the scheduling.  In fact, the
marginalization \eqref{eq:bap1} is reminiscent of a combined detector and VN
decoder for binary LDPC codes that are directly mapped to larger signal
constellations \cite{tenBrink2004}. Thanks to the symmetry of the $q$-SC, here
it is not necessary to actually sum over all $q$ symbol values.

\subsection{\qsc\ Front-end Approach}
\label{sec:front-end} 

The similarity of the bit-APP approach to combined detection and decoding
mentioned above points to a different view on the bit-symmetric coding scheme,
which is to consider a \qsc\ front-end for a binary LDPC decoder, similar to
approaches for iterative demapping and decoding \cite{Sanderovich2005,
Lechner2006}. These techniques involve proper iterations
between the demapper and the LDPC decoder, being treated as separate
functional blocks.

The \qsc\ front-end takes into account the correlation between the errors on
the $m$ bit layers. 
Its factor graph representation allows to formulate a message passing
algorithm that computes essentially the same quantities as the approach in
Sec.~\ref{sec:bit-APP}, but opens the way to EXIT analysis and displays more
clearly the opportunity for further complexity reduction by appropriate
message scheduling.

As before, let $\bv{x}$ denote the vector of $q$-ary channel input symbols
$\xj$ ($j=1,\ldots,n$) and let $\xji$ denote bit $i$ of symbol $j$. In the
same way, the output of the channel is represented by $\bv{y}$, $\yj$ and
$\yji$, and the errors are denoted by $\bv{e}$, $\ej$ and $\eji$,
respectively. We assume w.l.o.g.\ that $\eji=\xji\oplus\yji$, where $\oplus$
denotes addition in GF($2$), and $\ej=\xj\oplus\yj$, by extension.

Let $\chi_C(\bv{x})$ be the characteristic function of the code, which
evaluates to one if $\bv{x}$ is a codeword and to zero otherwise. Furthermore
assuming that the transmitted codewords are equally likely, the \emph{a
  posteriori} probability satisfies the proportionality relation
\begin{equation*}
  f_{\bv{X}|\bv{Y}}(\bv{x}|\bv{y}) \doteq \chi_C(\bv{x})
  f_{\bv{Y}|\bv{X}}(\bv{y}|\bv{x}). 
\end{equation*}
The $q$-ary channel is assumed to be memoryless (on the symbol level), 
leading to a factorization of $f_{\bv{Y}|\bv{X}}$ as
\begin{align*}
  f_{\bv{X}|\bv{Y}}(\bv{x}|\bv{y}) 	& \doteq \chi_C(\bv{x})
  \prod_{j=1}^n f_{Y|X}(\yj|\xj)\nonumber\\ 
  & = \chi_C(\bv{x}) \prod_{j=1}^n f_{Y,E|X}(\yj,\xj\oplus\yj|\xj),
\end{align*}
where we used the fact that $\xj$, $\ej$ and $\yj$ are
related in a deterministic way.

Given an error symbol $\ej$, the bit-layers are independent of each other,
leading to 
\begin{align*}
  f_{\bv{X}|\bv{Y}}(\bv{x}|\bv{y}) 	& \doteq \chi_C(\bv{x})
  \prod_{j=1}^n \Bigg\{ \ferr(e^j) f_{Y|X,E}(\yj|\xj,\ej)\Bigg\}\nonumber\\ 
  & = \chi_C(\bv{x}) \prod_{j=1}^n \left\{ \ferr(e^j) \prod_{i=1}^m
    \fchj(\yji|\xji,\eji) \right\}, 
\end{align*}
where $\fchj = f_{\Yji|\Xji,\Eji}$.

\begin{figure}
	\centering
	\psfrag{x1}[bc]{$x_1$}
	\psfrag{x2}[bc]{$x_2$}
	\psfrag{xm}[bc]{$x_m$}
	\psfrag{y1}[l]{$y_1$}
	\psfrag{y2}[l]{$y_2$}
	\psfrag{ym}[l]{$y_m$}
	\psfrag{e1}[bc]{$e_1$}
	\psfrag{e2}[bc]{$e_2$}
	\psfrag{em}[bc]{$e_m$}
	\psfrag{f1}[bc]{$\fch$}
	\psfrag{f2}[bc]{$\fch$}
	\psfrag{fm}[bc]{$\fch$}
	\psfrag{ferr}[bc]{$\ferr$}
	\psfrag{m}[bc]{$m$}
	\psfrag{efe}[br]{\scriptsize$\mu_{e_1 \to \ferr}(e_1)$}
	\psfrag{fee}[bl]{\scriptsize$\mu_{\ferr \to e_1}(e_1)$}
	\psfrag{fe}[bc]{\scriptsize$\mu_{\fch \to e_1}(e_1)$}
	\psfrag{ef}[bc]{\scriptsize$\mu_{e_1 \to \fch}(e_1)$}
	\psfrag{xf}[bc]{\scriptsize$\mu_{x_1 \to \fch}(x_1)$}
	\psfrag{fx}[bc]{\scriptsize$\mu_{\fch \to x_1}(x_1)$}
	\includegraphics[width=3.2in]{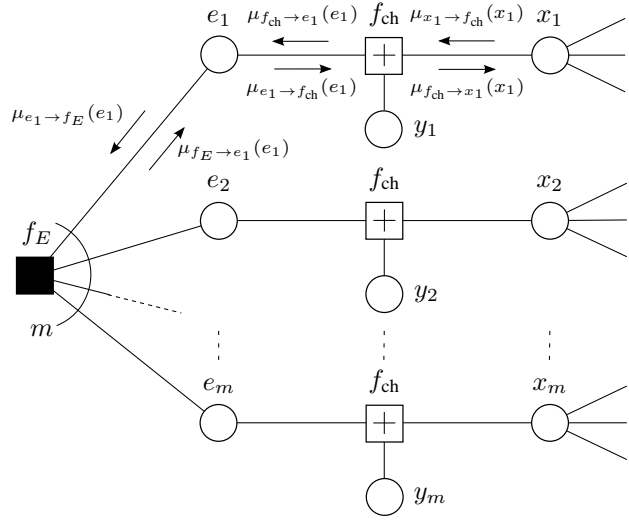}
	\caption{Factor graph representation of the $q$-SC front-end for
          symbol $\xj$ (the symbol index $j$ is omitted).} 
	\label{fig:graph}
\end{figure}

This factorization is shown for one symbol $\xj$ in Fig.~\ref{fig:graph},
where the edges on the right side are connected to the check nodes of the LDPC
code, i.e.\ the factorization of the characteristic function $\chi_C(\bv{x})$.
In the following, we will denote a message sent from a variable node $x$ to a
function node $f$ and vice versa as $\mu_{x \to f}(x)$ and $\mu_{f \to x}(x)$,
respectively \cite{Kschischang2001}.

\subsubsection{Message Passing Rules}
We apply the sum-product algorithm to compute the marginals
$f_{\Xji|\bv{Y}}(\xji|\bv{y})$ for all symbols $j$ and all bits $i$, i.e.\ to
perform \emph{a posteriori} decoding for every bit of the transmitted
codeword. In the following we will describe the operations for a single
$q$-ary symbol and therefore omit the symbol index $j$ for convenience.

First, we define the local functions and derive the message passing rules for
the function nodes $\fch$ and $\ferr$. Since the variable nodes $e_i$ have
degree two, they just forward the incoming messages, e.g.\ $\mu_{e_1 \to
  \fch}(e_1)=\mu_{\ferr \to e_1}(e_1)$ and $\mu_{e_1 \to \ferr}(e_1)=\mu_{\fch
  \to e_1}(e_1)$.

\paragraph{Bit Layer Channels}
In the binary sub-channels, $x_i$, $y_i$ and $e_i$ have to sum up to zero in
GF($2$), therefore the local functions $\fch$ are given by 
\begin{equation*}
	\fch(y_i|x_i,e_i) = [y_i = x_i \oplus e_i],
\end{equation*}
where $[P]$ evaluates to $1$ if the expression $P$ is true and to $0$
otherwise. 

Having defined the local function, we can derive the message passing algorithm
by applying the rules of the sum-product algorithm. Given a message
$\msg{x_i}{\fch}{x_i}$ and a received bit $y_i$, the channel function node
computes
\begin{align*}
  \msg{\fch}{e_i}{e_i}	& = \marg{e_i}\Bigg( \fch(y_i|x_i,e_i) \cdot
  \msg{x_i}{\fch}{x_i} \Bigg)\nonumber\\ 
  & = \msg{x_i}{\fch}{e_i \oplus y_i},
\end{align*}
where $\marg{a}$ denotes the marginalization over all variables except $a$. In
the same way, the message sent back to variable node $x_i$ is computed as 
\begin{align*}
  \msg{\fch}{x_i}{x_i}	& = \marg{x_i}\Bigg( \fch(y_i|x_i,e_i) \cdot
  \msg{e_i}{\fch}{e_i} \Bigg)\nonumber\\ 
  & = \msg{e_i}{\fch}{x_i \oplus y_i}.
\end{align*}

\paragraph{Error Patterns}
The function node $\ferr$ represents the probability that a certain binary
error pattern occurs. In the $q$-ary symmetric channel, every error pattern
has the same probability $\frac{\eps}{q-1}$, except the all-zero pattern that
has probability $1-\eps$, thus
\begin{equation*}
  \ferr(e_1,\ldots,e_m) =
  \begin{cases}
    1-\eps, & \text{if } e_1=\ldots=e_m=0\\
    \frac{\eps}{q-1},	& \text{otherwise.}
  \end{cases}
\end{equation*}
The derivation of the outgoing messages of $\ferr$ leads to
\begin{align*}
  \msg{\ferr}{e_i}{e_i}	& = \marg{e_i}\Bigg(\ferr(e_1,\ldots,e_m) \cdot
  \prod_{i'\neq i} \msg{e_{i'}}{\ferr}{e_{i'}} \Bigg)\nonumber\\ 
  & = \left\{
    \begin{array}{ll}
      (1-p) \cdot \beta_{[i]} + \frac{\eps}{q-1}\left(1-\beta_{[i]} \right) &
      \mbox{if $e_i=0$}\\
      \frac{\eps}{q-1} 				
      & \mbox{if $e_i=1$},
    \end{array}
  \right.
\end{align*}
where $\beta_{[i]}$ is defined as
\begin{align}
  \label{eq:betai}
  \beta_{[i]} = \prod_{i' \neq i} \msg{e_{i'}}{\ferr}{0}.
\end{align}
Appendix~\ref{sec:msg_simplification} outlines the simplification of the
different messages using scalar quantities.

\subsubsection{Complexity}

Clearly, the processing complexity of the front-end is dominated by
\eqref{eq:betai} and is thus $O(\log q)$ per symbol, as in verification
decoding \cite{Luby2005}. This means that the overall decoding complexity
scales linearly in the number of transmitted bits, independently of the symbol
alphabet size $q$. Complexity may be further reduced by a constant factor
(i.e.\ \emph{not} its order) if the front-end messages are not recomputed on
every LDPC decoder iteration; this is often sufficient in practice.

\subsection{EXIT Analysis of the \qsc\ Front-end}
\label{sec:exit}

In this section, we analyze the \qsc\ front-end using extrinsic information
transfer (EXIT) charts \cite{Ashikhmin2004b}, which will later also be used
for the design (optimization) of LDPC codes.

Let the \emph{a priori} and \emph{extrinsic} messages for the front-end denote
the messages between the bit layer sub-channels $\fch$ and the bit nodes $x_i$
of the LDPC code, that is, $\mu_{x_i \to \fch}(x_i)$ and $\mu_{\fch \to
  x_i}(x_i)$, respectively. For the EXIT chart analysis, the front-end is
characterized by the transfer function $I_{e}(I_{a})$, where $I_{a}$ and
$I_{e}$ denote the \emph{a priori} and the \emph{extrinsic} mutual
information, respectively, between the messages and the corresponding bits.

First, we derive the EXIT function of the front-end for the case when the
\emph{a priori} messages are modeled as coming from a binary erasure channel
(BEC). In this special case, the EXIT functions have important properties
(e.g.\ the area-property), which were shown in \cite{Ashikhmin2004b}.

In the case of a BEC, the variables $x_i$ are either known to be error-free or
erased, thus simplifying the computation of $\beta_{[i]}$ defined in
\eqref{eq:betai} as follows. The product in \eqref{eq:betai} has $m-1$ factors,
taking on the value $1/2$ if the bit $x_i$ is erased, $0$ if the bit $x_i$ is
not in agreement with the received bit $y_i$ and $1$ otherwise. If at least
one message is not in agreement with the received bit, a symbol error occurred
and $\beta_{[i]}=0$. Otherwise, $\beta_{[i]}=2^{-t}$, where $t$ denotes the
number of erased messages.

\begin{figure}[t]
  \centering
  \includegraphics[width=3.2in]{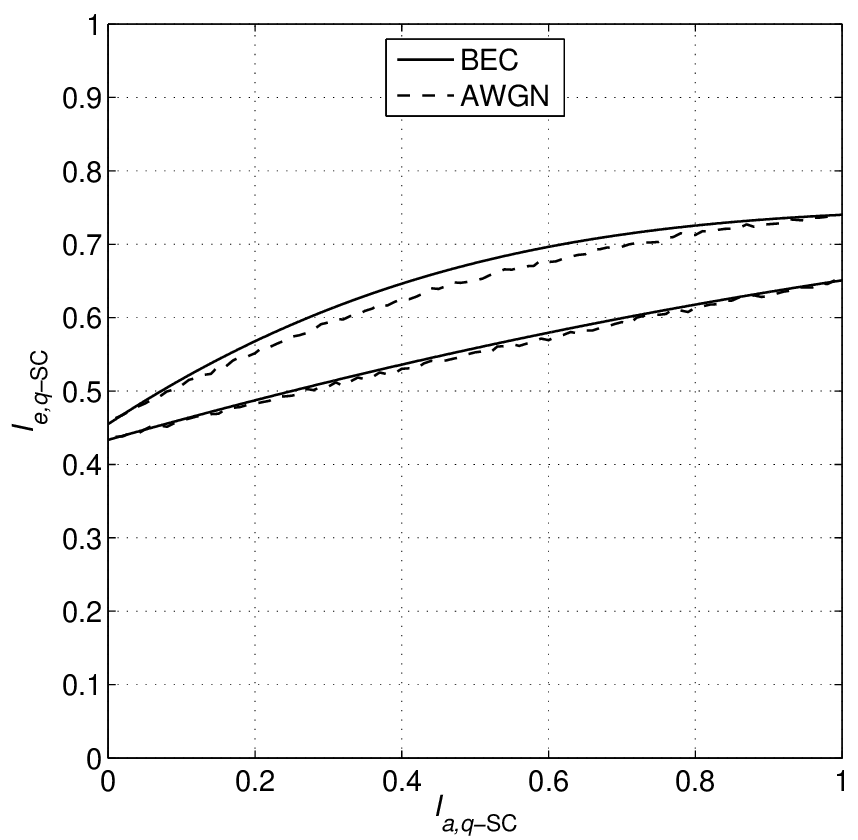}
  \caption{EXIT function for the $q$-SC front-end with BEC (analytical) and
    Gaussian (simulated) \emph{a priori} messages for $\epsilon=0.25$, 
    $m=4$ (lower curves) and $m=8$ (upper curves).}
  \label{fig:exitbec}
\end{figure}

Since $\beta_{[i]}=0$ corresponds to $\epsilon_{i}=1/2$ through
\eqref{eq:epsiloni}, this case is equivalently described by an
erasure. Therefore, the extrinsic messages can be modeled as being transmitted
over a binary symmetric erasure channel (BSEC), with parameters
$\epsilon_{m-t}$ and $\delta_{m-t}$ given by \eqref{eq:eps} and
\eqref{eq:delta}, respectively, since $t$ erasures correspond to the situation
in layer $i=m-t$ in the layered scheme of Section~\ref{sec:layer_scheme}. Its
capacity is thus
\begin{equation}
	\label{eq:bseccap}
	I_t = \left(1-\delta_{m-t}\right)
        \left(1-h\left(\frac{\epsilon_{m-t}}{1-\delta_{m-t}}\right)\right). 
\end{equation}
In order to compute the extrinsic information $I_e(I_a)$ at the output of the
front-end as
\begin{equation}
	\label{eq:exitbec}
	I_e(I_a) = \sum_{t=0}^{m-1} I_t \lambda_t(I_a),
\end{equation}
we have to compute the probability $\lambda_t(I_a)$ that $t$ messages are
erased given \emph{a priori} mutual information $I_a$. Let $\Delta_a=1-I_a$
denote the probability that an \emph{a priori} message is erased. The
probability that $t=0,\ldots,m-1$ out of $m-1$ messages are erased is computed
as
\begin{align}
	\label{eq:probt}
	\lambda_t(I_a) 	&= \binom{m-1}{t} \Delta_a^{t}
        \left(1-\Delta_a\right)^{m-1-t}\nonumber\\ 
	&= \binom{m-1}{t} \left(1-I_a\right)^{t} I_a^{m-1-t}.
\end{align}
The EXIT function for BEC \emph{a priori} messages can therefore be computed
using \eqref{eq:eps}, \eqref{eq:delta}, \eqref{eq:bseccap} and
\eqref{eq:probt} in \eqref{eq:exitbec}.

\begin{Thm}
  An iterative decoder for the bit-symmetric scheme using a $q$-SC front-end
  can attain $q$-SC capacity.
\end{Thm}

\begin{IEEEproof}
  According to the area theorem for EXIT charts \cite{Ashikhmin2004b}, the
  capacity of each bit layer is given by the area below the EXIT function, if
  the \emph{a priori} messages are modeled as coming from a BEC. Therefore,
  the capacity of the overall channel is given by
\begin{eqnarray}
  \label{equ:area}
  C_{\text{tot}} & = & m \int_0^1 \sum_{t=0}^{m-1} I_t \lambda_t(I_a) \ud I_a
  \nonumber\\  
  & = & m \sum_{t=0}^{m-1} I_t \int_0^1 \lambda_t(I_a) \ud I_a\nonumber\\
  & = & m \sum_{t=0}^{m-1} I_t \binom{m-1}{t} \int_0^1 \left(1-I_a\right)^{t}
  I_a^{m-1-t} \ud I_a\nonumber\\
  & = & m \sum_{t=0}^{m-1} I_t \frac{(m-1)!}{(m-1-t)!t!}
  \frac{t!(m-1-t)!}{m!}\nonumber\\ 
  & = &  \sum_{t=0}^{m-1} I_t = \cqsc,
\end{eqnarray}
where we used the fact that the integral corresponds to the definition of the
beta function. The last sum in \eqref{equ:area} is indeed equal to the
capacity of the \qsc, as was already shown in the proof of Theorem~1.
\end{IEEEproof}

Without iterative processing, the maximum achievable rate is given by
$I_e(0)$. In that case, all \emph{a priori} messages are erased and one has 
\begin{equation*}
  I_{m-1} = 1-h\left(\epsilon_{1}\right)
        = 1-h\left(\frac{pq}{2(q-1)}\right)
        = \cbsc,
\end{equation*}
which corresponds to the simplistic coding approach where the \qsc\ is
decomposed into $m$ BSCs. 

Figure~\ref{fig:exitbec} shows an EXIT chart for BEC \emph{a priori} messages.
If the \qsc~would be decomposed into BSCs (without iterating between the
front-end and the LDPC code and without using the layered scheme) then the
value at $I_{a}=0$ corresponds to the capacity of these BSCs. In contrast, the
area below the EXIT functions corresponds to the normalized capacity of the
\qsc.
When using this front-end with an LDPC code, the
\emph{a priori} messages can not be modeled as coming from a BEC channel. For
code design, we will make the approximation that the messages are Gaussian
distributed, which is a common assumption when using EXIT charts. The
simulated EXIT function using Gaussian \emph{a priori} messages is also shown
in Figure~\ref{fig:exitbec}.

\section{LDPC Design and Construction}
\label{sec:design}

After obtaining the EXIT function of the \qsc\ front-end, we can design an
LDPC code. Code design is an optimization problem that selects the degree
distributions of the code in order to maximize the rate, under the constraint
that the EXIT function of the overall LDPC code does not intersect the EXIT
function of the \qsc\ front-end.

Joint optimization of the variable and check
node degree distributions is a nonlinear problem, but it was shown in
\cite{Lechner2006} that the optimization of the check node degree distribution
given a fixed variable node degree distribution is a linear programming
problem, which can be solved efficiently.

\begin{figure}[t]
  \centering
  \includegraphics[width=3.2in]{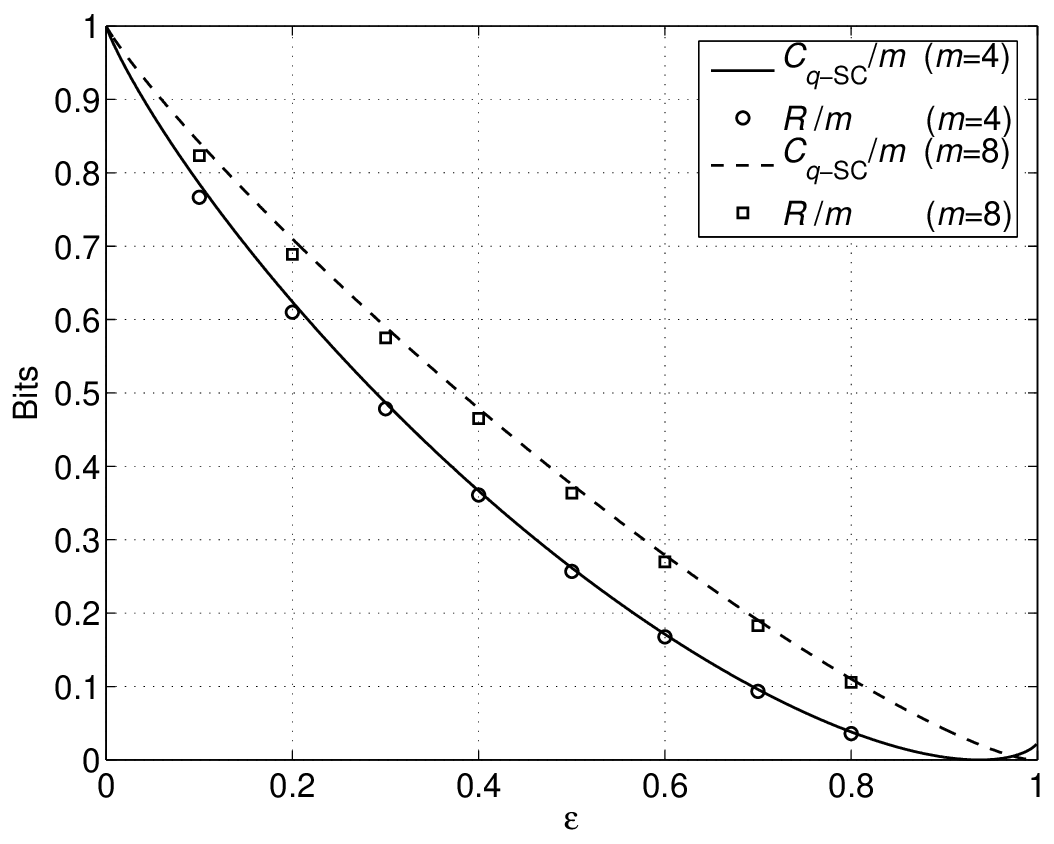}
  \caption{Rate of optimized codes versus normalized capacity of the \qsc\ for
    $m=4$ and $m=8$.} 
  \label{fig:rates}
\end{figure}

\begin{figure}[ht]
  \centering
  \includegraphics[width=3.2in]{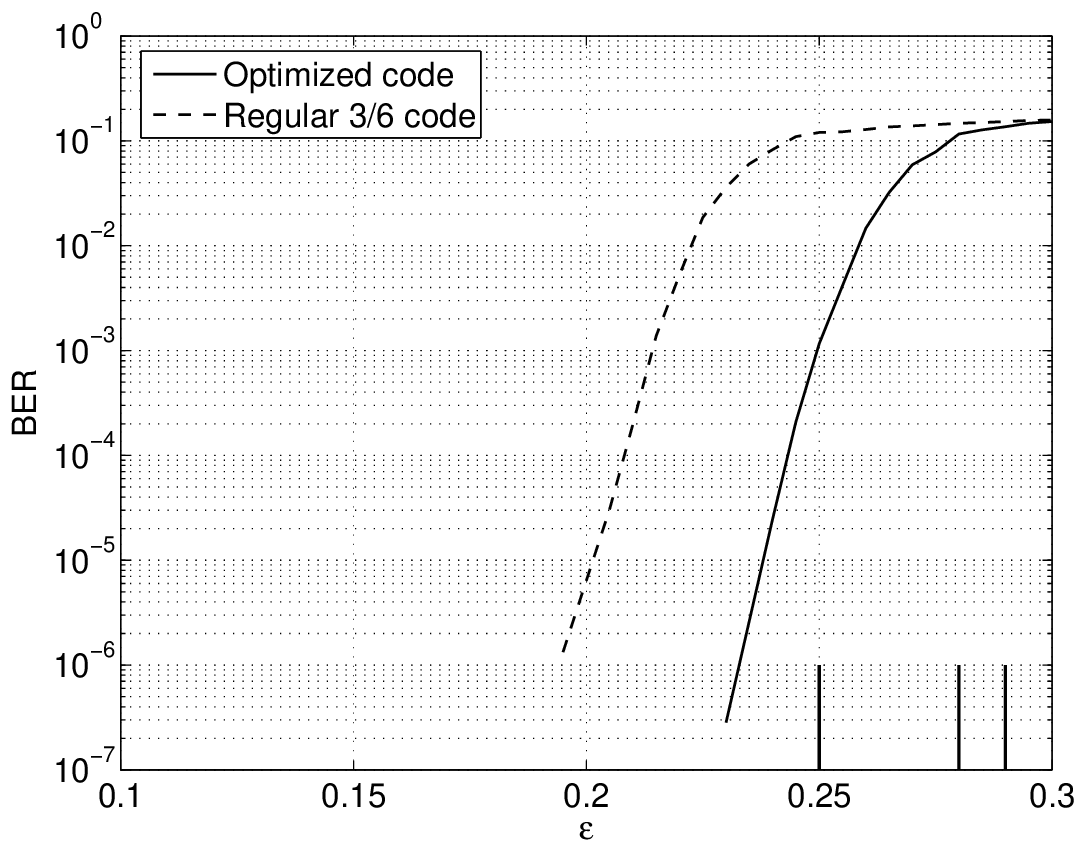}
  \caption{Bit error rate simulation for rate 1/2 LDPC decoding ($m=4$,
    $k=1500$ symbols). Marks indicate decoding threshold for regular code
    ($\epsilon=0.25$), optimized code ($\epsilon=0.28$) and Shannon limit
    ($\epsilon=0.29$).}
  \label{fig:n12000}
\end{figure}

To reduce the optimization complexity and to obtain practical degree
distributions, we choose three non-zero variable node degrees (compare e.g.,
the approach in \cite{tenBrink2003}). This allows to perform an exhaustive
search over the variable node degree distribution, and for each variable node
degree distribution we optimize the check node degree distribution as
described in \cite{Lechner2006} with a maximum check node degree
$d_{c,max}=50$.

Fig.~\ref{fig:rates} shows the normalized capacity of the \qsc\ and the
obtained optimized code rates (under the Gaussian approximation) versus the
error probability $\epsilon$ of the \qsc\ for $m=4$ and $m=8$.  Using this
code optimization, we are able to design codes that perform close to capacity
over a wide range of error probabilities for moderate $q$.

For the actual code construction, we used a modified version of the PEG
algorithm \cite{Hu2005,PEGSoftware}, which allows us to construct codes with a
specific variable and check node degree distribution.  To avoid (short) cycles
in the overall factor graph, one has to ensure that the bits of a given
$q$-ary symbol do not participate in the same parity check. This can be
achieved by using an appropriate interleaver between the \qsc\ front-end and
the LDPC decoder. Another way would be to use the multi-edge type framework
for LDPC code where such constraints can be explicitly specified
\cite[Chap.~7]{MCT2007}.

To verify our derivations, we performed bit error rate simulations for an
optimized binary LDPC code of rate $R=1/2$ and length $N=12000$, shown in
Fig.~\ref{fig:n12000}. This code is optimized for a \qsc\ with $m=4$, thus
$N=12000$ bits correspond to $k=1500$ $q$-ary symbols.  It has a decoding
threshold of $\epsilon=0.28$ (the Shannon limit for $R=1/2$ is at
$\epsilon=0.29$). As a comparison we also show the bit error rate for a
regular LDPC code with $d_v=3$ and $d_c=6$ which has a decoding threshold of
$\epsilon=0.25$.

\section{Generalization to Conditionally Independent Binary Sub-channels}
\label{sec:gen_scheme}

The decoder in Section~\ref{sec:sym_scheme} is seen to easily extend to
$q$-ary channels with modulo-additive noise, where the $m$ bits of the noise
(error pattern) are independent if an error occurs, i.e.\ if at least
one bit is nonzero. This assumption leads to define the $q$-ary symmetric
channel with conditionally independent binary sub-channels (\qscs), with
transition probabilities
\begin{align*}
  P_{Y|X}(y|x) &= 
  \begin{cases}
    1-\eps, & y=x,\\
    \alpha \prod_{i=1}^m \epsilon_{i|*}^{x_i\oplus y_i}
\left(1-\epsilon_{i|*}\right)^{1\oplus x_i\oplus y_i}, & y \neq x,
  \end{cases}\\
\text{where}\quad
\alpha &= \frac{\eps}{1-\prod_{i=1}^m\left(1-\epsilon_{i|*}\right)}\nonumber
\end{align*}
and $\oplus$ denotes GF($2$) 
addition (this definition encompasses all channels satisfying the above
conditional independence assumption, from which it may be derived
axiomatically). The defining parameters are the symbol error probability
$\eps$ and the $m$ conditional probabilities $\epsilon_{i|*}$, which are the
bit error probabilities conditioned on the event that at least one of the
other $m-1$ bits is in error. The marginal bit error probabilities will be
$\epsilon_{i}=\alpha\epsilon_{i|*}$.  Since the \qscs\ is strongly symmetric,
the uniform input distribution achieves its capacity
\begin{multline*}
  \ochf
  \cqscs = m - \alpha \sum_{i=1}^m h(\epsilon_{i|*}) + (1-\eps)\log_2(1-\eps)
  \tcbr 
  +\alpha\log_2\alpha - (\alpha-\eps)\log_2(\alpha-\eps).
  \ochf
\end{multline*}
When $\eps$ is such that
$\alpha=1$, the \qscs\ reduces to $m$ independent BSCs, while for
$\epsilon_{i|*}=\half$, $i=1,\ldots, m$, it becomes an ordinary \qsc.

\begin{figure*}[!t]
\normalsize
\begin{align}
\label{eq:appllr_gen}
  L_{\mathrm{app}}(\Xji\!=\!\yji) &= \log\left( 
  \frac{(1-\eps)\beta^{j} + \alpha (1-\epsilon_{i|*})\pji\left[
\prod_{k\neq i}\left(\epsilon_{k|*} + \pjk 
- 2\epsilon_{k|*} \pjk\right) 
-\prod_{k\neq i}(1-\epsilon_{k|*})\pjk \right]}
{\alpha \epsilon_{i|*} (1-\pji)
\prod_{k\neq i}\left(\epsilon_{k|*} + \pjk - 2\epsilon_{k|*} \pjk\right)}
  \right) \nonumber\\
 &= \log\left( 
\frac{(1-\alpha)\beta^j_{[i]} + \alpha (1-\epsilon_{i|*})
\prod_{k\neq i}\left(\epsilon_{k|*} + \pjk 
- 2\epsilon_{k|*} \pjk\right)}
{\alpha \epsilon_{i|*}    
\prod_{k\neq i}\left(\epsilon_{k|*} + \pjk - 2\epsilon_{k|*}
  \pjk\right)} \right)
  +\log\left(\frac{\pji}{1-\pji} 
  \right) \nonumber\\
  &= \underbrace{\log\left( \frac{1-\epsilon_{i|*}}{\epsilon_{i|*}} 
+ \frac{(1-\alpha)\beta^j_{[i]}}
{\alpha \epsilon_{i|*}    
\prod_{k\neq i}\left(\epsilon_{k|*} + \pjk - 2\epsilon_{k|*}
  \pjk\right)}
\right)}_{\text{\normalsize $\lchqsc(\Xji\!=\!\yji)$}}
+ L_{\mathrm{extr}}(\Xji\!=\!\yji).
\end{align}
\hrulefill
\vspace*{4pt}
\end{figure*}

The unnormalized bit APP (compare \eqref{eq:bap2}) is obtained as
\begin{align*}
\ifthenelse{\boolean{CLASSOPTIONtwocolumn}}{&}{}%
P(\Xji=\xji|\bv{Y}=\bv{y}) 
\ifthenelse{\boolean{CLASSOPTIONtwocolumn}}{\\ &\qquad}{}%
\doteq  \begin{cases}
  (1-\eps)\prod_{k=1}^m \pjk\\ \quad + \alpha (1-\epsilon_{i|*})\pji\Big[
\prod_{k\neq i}\big(\epsilon_{k|*} + \pjk 
- 2\epsilon_{k|*} \pjk\big) 
\ifthenelse{\boolean{CLASSOPTIONtwocolumn}}{\\ \qquad}{} 
-\prod_{k\neq i}(1-\epsilon_{k|*})\pjk
\Big],\ifthenelse{\boolean{CLASSOPTIONtwocolumn}}{\hfill}{\quad}
      \xji = \yji,\\
      \alpha \epsilon_{i|*} (1-\pji)
\prod_{k\neq i}\big(\epsilon_{k|*} + \pjk - 2\epsilon_{k|*} \pjk\big),
\ifthenelse{\boolean{CLASSOPTIONtwocolumn}}{\\}{}  \hfill \xji \neq \yji,
  \end{cases} 
\end{align*}
leading to the LLR expression \eqref{eq:appllr_gen}. The quantities $\pji$,
$\beta^j$ and $\beta^j_{[i]}$ are as defined in Section~\ref{sec:sym_scheme}.
Like in the special case of the \qsc, for $\beta^j_{[i]} \to 0$ the channel
LLR $\lchqsc$ tends to a constant, which is equal to the LLR of the binary
sub-channel conditioned on the occurrence of an error on one or more of the
other sub-channels. The initial value of $\lchqsc$ can again be obtained by
assuming a worst-case of $\pjk=\half$, $\beta^j_{[i]}=2^{-m+1}$ in
\eqref{eq:appllr_gen}, yielding
\begin{align*}
\lchqsc^{(0)}(\Xji\!=\!\yji) &= 
\log\left( \frac{1-\epsilon_{i|*}}{\epsilon_{i|*}} +
  \frac{1-\alpha}{\alpha\epsilon_{i|*}} \right)\\ &=
\log\left( \frac{1-\alpha\epsilon_{i|*}}{\alpha\epsilon_{i|*}}\right)
= \log\left( \frac{1-\epsilon_{i}}{\epsilon_{i}}\right),
\end{align*}
which is the channel LLR of the marginal BSC.

\appendix
\subsection{Simplification of the Messages}
\label{sec:msg_simplification}

Since all involved variables are binary, we can represent the messages by
scalar quantities. For the messages from and to the variable nodes $\xji$ of
the LDPC code, we use log-likelihood ratios of the corresponding messages 
\begin{equation*}
  L_{a,i} = \log\frac{\msg{x_i}{\fch}{0}}{\msg{x_i}{\fch}{1}},
\end{equation*}
\begin{equation*}
  L_{\text{ch},i} = \log\frac{\msg{\fch}{x_i}{0}}{\msg{\fch}{x_i}{1}}.
\end{equation*}

The messages from the function nodes $\fch$ to the node $\ferr$ are defined as
the probability of no error
\begin{equation*}
  p_i = \msg{\fch}{e_i}{0} = \msg{e_i}{\ferr}{0},
\end{equation*}
and the messages from the function node $\ferr$ are defined as the error
probability 
\begin{equation*}
  \epsilon_i = \frac{\msg{\ferr}{e_i}{1}}{\msg{\ferr}{e_i}{0} +
    \msg{\ferr}{e_i}{1}}. 
\end{equation*}

Using these definitions, we can reformulate the computation rules at the
function nodes as 
\begin{equation*}
  p_i = \frac{e^{\frac{L_{a,i}}{2}\cdot(1-2y_i)}}{e^{\frac{L_{a,i}}{2}} +
    e^{-\frac{L_{a,i}}{2}}}, 
\end{equation*}
and
\begin{align}
  \label{eq:epsiloni}
  \epsilon_i	& = \frac{\frac{\eps}{q-1}}{(1-\eps) \cdot \beta_{[i]} +
    \frac{\eps}{q-1} \left(1 - \beta_{[i]} \right) +
    \frac{\eps}{q-1}}\nonumber\\  
  & = \frac{\eps}{2\eps + \beta_{[i]} (q - \eps q - 1)}
\end{align}

For the initialization, we set all a-priori L-values $L_{a,i}$ to zero,
leading to $\beta_{[i]}=2^{-(m-1)}=2/q$. Inserting this in \eqref{eq:epsiloni}
leads to
\begin{equation*}
	\label{eq:epsiloninit}
	\epsilon_{i} = \frac{\eps q}{2(q-1)} = \epsbsc.
\end{equation*}

Finally, the messages sent from the function nodes $\fch$ to the variable
nodes $x_i$ are computed as 
\begin{equation*}
  L_{\text{ch},i} = (1-2y_i) \cdot \log\frac{1-\epsilon_i}{\epsilon_i}.
\end{equation*}

\section*{Acknowledgments}
The authors would like to thank Jossy Sayir for helpful discussions and for
pointing out the similarity with iterative demapping and decoding.

\bibliographystyle{IEEEtran}
\bibliography{cw_jabref}  

\end{document}